\documentclass[12pt,preprint]{aastex}

\shorttitle{Mass segregation in globular clusters}
\shortauthors{Baumgardt et al.}

\begin{document}


\title{Evidence for primordial mass segregation in globular clusters}


\author{Holger Baumgardt\altaffilmark{1},        
        Guido De Marchi\altaffilmark{2},
        Pavel Kroupa\altaffilmark{1}}

\altaffiltext{1}{
        Argelander-Institute for Astronomy, University of Bonn, 
        Auf dem H\"ugel 71, 53121 Bonn, Germany}

\altaffiltext{2}{
        European Space Agency, Space Science Department, Noordwijk, Netherlands}


\begin{abstract}

We have studied the dissolution of initially mass segregated and unsegregated star clusters 
due to two-body relaxation in external tidal fields,
using Aarseth's collisional $N$-body code NBODY4 on GRAPE6 special-purpose computers.
When extrapolating results of initially not mass segregated models to globular clusters, we
obtain a correlation between the time until destruction and the slope of the mass function, in
the sense that globular clusters which are closer to dissolution are more strongly
depleted in low-mass stars. This correlation fits observed mass functions of most globular
clusters. The mass functions of several globular clusters are however
more strongly depleted in low-mass stars than suggested by these models.
Such strongly depleted mass functions can be explained if globular clusters started
initially mass segregated. Primordial mass segregation also explains the correlation
between the slope of the stellar mass function and the cluster concentration which was recently
discovered by De Marchi et al.\ (2007). In this case, it is possible that all globular clusters
started with a mass function similar to that seen in young open clusters in the present-day
universe, at least for stars below $m=0.8$ M$_\odot$. This argues for a near universality of
the mass function for different star formation environments and metallicities in the range
$-2 < \mbox{[Fe/H]} < 0$.  We finally describe a novel algorithm which can initialise stationary 
mass segregated clusters with arbitrary density profile and amount of mass segregation.
\end{abstract}


\keywords{galaxies: star clusters---methods: N-body simulations---stellar dynamics---stars: luminosity function, mass function}


\newcommand{\msun}{M_{\odot}}
\def\apgt{\ {\raise-.5ex\hbox{$\buildrel>\over\sim$}}\ }
\def\aplt{\ {\raise-.5ex\hbox{$\buildrel<\over\sim$}}\ }

\section{Introduction}

In recent years, stellar mass functions have been obtained for an increasing number of globular clusters 
by deep {\it HST} and {\it VLT} measurements (see DeMarchi et al.\ 2007 and references therein). These observations 
have shown that there
is a considerable spread in the present-day mass functions of individual clusters, and that a number
of star clusters are strongly depleted in low-mass stars. If one expresses the mass function
of a cluster as a power-law\footnote{Note that \citet{dpp07} used $dN/dm \sim m^{\alpha}$ in their paper.}
by $dN/dm \sim m^{-\alpha}$, where $N$ is the number of stars per
unit mass $m$, the observed slopes range from between $\alpha=1.9$ to $\alpha=-0.9$ for stars with
masses in the range $0.3 < m < 0.8$ M$_\odot$. For clusters 
where information from different radii is available, the data point to a global decrease of 
the number of low-mass stars in the clusters, rather than a local effect due to mass segregation. 

The depletion of low-mass stars can in principle be understood by mass segregation and the preferential 
loss of low-mass stars as a result of the dynamical evolution of star clusters. Indeed, using
direct $N$-body simulations, \citet{bm03} found a correlation between the observed and expected slopes 
for the then available sample of star clusters. However, for a number of clusters, the difference
between theoretical and expected slope is far too large to be explained just by observational uncertainties.

This is emphasised by \citet{dpp07}, who found a correlation between the mass function slope $\alpha$
and the concentration parameter $c=\log_{10}(r_t/r_c)$ for globular clusters, where $r_t$ and $r_c$ are the tidal
and core radius of the cluster as determined from the projected light density profile. The correlation
found by \citet{dpp07} is in the sense that clusters with small values of $c$ are depleted in low-mass 
stars, while clusters with large values of $c$ have mass functions still rising towards small masses. Since 
simulations show that
mass segregation and the preferential loss of low-mass stars should only happen after a cluster has gone 
into core-collapse, and since core-collapse is connected to the shrinkage of the core size $r_c$,
the observed correlation is the exact opposite of the theoretically expected one.

One possible interpretation of this finding would be that star clusters that formed more concentrated have a
bottom-heavy IMF, which would be a challenge to star formation theories and dispose the universality of the
IMF. However this conclusion needs to be tested by taking into account the stellar-dynamical evolution of
the clusters. 

In the present paper we compare the observational results with theoretical predictions by \citet{bm03} 
(BM03), who have performed a large parameter study of initially not mass segregated multi-mass clusters 
evolving under the combined influence of 
relaxation, stellar evolution and an external tidal field. We also report results of new simulations
of multi-mass star clusters which start initially mass segregated. Initial mass segregation is 
expected to occur in star clusters as a result of star formation feedback in dense gas clouds \citep{mur96} 
and due to competitive gas accretion and mutual mergers between protostars \citep{bon02}. Numerous
studies have also found observational evidence for it in young star clusters of the Milky Way and the Magellanic
Clouds \citep{bon98,getal04,cdz07}, so that it is certainly possible that globular clusters started
mass segregated.

The paper is organised as follows: In \S2 we compare the results of simulations of non-mass-segregated
clusters done by \citet{bm03} with the observed mass function slopes of globular clusters. In \S3 we describe 
the numerical simulations of star clusters with primordial mass segregation and in \S4 we compare the results 
of these runs with the observations. We briefly summarize in \S5.

\section{Models without primordial mass segregation}

BM03 performed a large set of $N$-body simulations of multi-mass star clusters moving in external tidal 
fields and evolving under the combined influence of two-body relaxation, an external tidal field and stellar 
evolution.
All models contained between 8.192 to 131.072 stars and started with a {\it canonical} mass function that
consisted of two power-laws with slope $\alpha=1.3$ for stars between 0.08 and 0.5 M$_\odot$ and slope
$\alpha=2.3$ for more massive stars \citep{k01}.
The clusters moved on circular or eccentric orbits through an isothermal galaxy with circular velocity 
$V_C=220$ km/sec. BM03 obtained the following expression for the lifetime $T_{Diss}$ of a star cluster:
\begin{equation}
\frac{T_{Diss}}{\mbox{[Myr]}} \; = \; \beta \; \left( \frac{N}{ln(\gamma \, N)} \right)^x \; 
  \frac{R_G}{\mbox{[kpc]}} \; \left( \frac{V_G}{220\; \mbox{km/sec}} \right)^{-1} (1-\epsilon)\,\, ,
\label{eltime}
\end{equation}
where $N$ is the number of cluster stars, $\gamma=0.02$ a constant in the Coulomb logarithm and 
$R_G$ and $\epsilon$ the apocenter distance and eccentricity of the cluster orbit, respectively.
The constants $\beta$
and $x$ were found to depend on the density profile. For King $W_0=7$ models, $x$ and $\beta$ are given 
by $x=0.82$ and $\beta = 1.03$. BM03 found that mass is lost more or less linearly with time 
from a star cluster, except for the mass lost due to stellar evolution,
which decreases the initial mass by about 30\% within the first Gyr. The mass left at a
time $T<T_{Diss}$ can therefore be approximated by
\begin{equation}
M(T) = 0.70 \, M_0 \, (1-T/T_{Diss}) \;\;\;\;\; .
\label{emfunc}
\end{equation}

BM03 also found that, while the clusters are dissolving, mass segregation causes massive stars to sink into the 
cluster center and low-mass stars to move to the outer parts, where they are easily removed by the tidal field,
so that the global mass function of stars gets depleted in low-mass stars. By fitting power-law
mass functions $dN/dm \sim m^{-\alpha}$ to the mass function of stars with $m< 0.5$ M$_\odot$, BM03 
derived the following expression for the change in the slope of the mass function:
\begin{equation}
 \alpha = 1.3-1.51 \left(\frac{T}{T_{Diss}}\right)^2 + 1.69 \left(\frac{T}{T_{Diss}}\right)^3 -
  1.50 \left(\frac{T}{T_{Diss}}\right)^4 \;.
\end{equation}
They found that this expression gave a good fit to the change of the mass function for a wide range of 
initial cluster orbits and cluster masses. Since observed mass function slopes of the clusters in \citet{dpp07} are 
determined mainly from stars with masses $0.3$ M$_\odot < m < 0.8$ M$_\odot$, we have re-analysed the data by BM03 and
find that the following formula fits the change of the mass function in this range:
\begin{equation}
 \alpha = 1.74-0.34 \frac{T}{T_{Diss}} 
 + 4.52 \left(\frac{T}{T_{Diss}}\right)^2 - 7.59 \left(\frac{T}{T_{Diss}}\right)^3 +
  5.86 \left(\frac{T}{T_{Diss}}\right)^4 \;.
\label{mftf}
\end{equation}
The runs by BM03 also indicated that the mass-to-light ratio drops as a cluster evolves and loses preferentially low-mass stars
which do not contribute much to the overall cluster light. The results of BM03 (their Fig.\ 14) can be fitted 
by the relation
\begin{equation}
 M/L = 1.5-0.5 \frac{T}{T_{Diss}} \;.
\label{mlr}
\end{equation}

 Using the above equations \ref{eltime}, \ref{emfunc} and \ref{mlr}, we can calculate the initial mass of individual globular 
clusters, provided their orbits and present-day luminosities are known. One way to do this is to first guess two initial masses 
$M_{Low}$ and $M_{Up}$ which lead to too small and 
too large present-day masses, and then iterate to the correct initial mass 
by interval-halving.  
Once the initial masses and dissolution times are known, the expected present-day mass function slopes of low-mass stars can be 
calculated from eq. \ref{mftf}. Table~1 
and Fig.~1 compare our predictions with the observed slopes from \citet{dpp07}. We have taken the 
pericenter and apocenter distances from \citet{di99}, except for NGC 6496 for which a circular orbit
at the current Galactocentric distance was assumed since its proper motion is not known.
The integrated luminosities were taken from \cite{h96}. We assumed an age of $T=12$ Gyr for the Galactic globular cluster system.

\begin{table*}[t]
\caption[]{Observed and theoretical mass function slopes for globular clusters. The first three columns give the name,
observed global mass function slope $\alpha$ and cluster concentration $c$, taken from \citet{dpp07}. The next 
columns give the absolute luminosity, pericenter and apocenter distance taken from \citet{h96} and \citet{di99}. The 
final columns come from our fit using eqs.\ 1, 2, 4 and 5, and give the lifetime, initial and present-day cluster mass and expected mass function slope for stars in the
range $0.3 < m < 0.8$ M$_\odot$ of initially unsegregated clusters.\\}
\begin{tabular}{lrcrrrrccc}
 \multicolumn{1}{c}{Cluster} &$\alpha_{Obs}$ &$c$ & \multicolumn{1}{c}{$M_V$} &$R_{Peri}$&$R_{Apo}$ & $T_{Diss}$ & $M_{C ini}$ & $M_{C}$ & $\alpha_{Th}$  \\
 &  & & & [kpc] & [kpc] & [Gyr] & [$M_\odot$]  & [$M_\odot$] & \\ 
NGC   104 &  1.2 &  2.03 & -9.42 &  5.2 &   7.3 &   85.7 & $1.1 \cdot 10^6$ & $7.0 \cdot 10^5$ &  1.73 \\
NGC   288 &  0.0 &  0.96 & -6.74 &  1.7 &  11.2 &   17.4 & $2.2 \cdot 10^5$ & $4.8 \cdot 10^4$ &  0.98 \\
NGC  2298 & -0.5 &  1.28 & -6.30 &  2.3 &  15.7 &   17.7 & $1.4 \cdot 10^5$ & $3.2 \cdot 10^4$ &  1.02 \\
Pal     5 &  0.4 &  0.70 & -5.17 &  7.0 &  19.0 &   14.3 & $1.0 \cdot 10^5$ & $1.0 \cdot 10^4$ &  0.35 \\
NGC  5139 &  1.2 &  1.61 &-10.29 &  1.2 &   6.2 &   55.7 & $2.7 \cdot 10^6$ & $1.5 \cdot 10^6$ &  1.69 \\
NGC  5272 &  1.3 &  1.84 & -8.93 &  5.5 &  13.4 &   82.6 & $7.2 \cdot 10^5$ & $4.5 \cdot 10^5$ &  1.73 \\
NGC  6121 &  1.0 &  1.59 & -7.20 &  0.6 &   5.9 &   14.3 & $7.0 \cdot 10^5$ & $6.7 \cdot 10^4$ &  0.30 \\
NGC  6218 & -0.1 &  1.29 & -7.32 &  2.6 &   5.3 &   21.3 & $2.6 \cdot 10^5$ & $8.7 \cdot 10^4$ &  1.33 \\
NGC  6254 &  1.1 &  1.40 & -7.48 &  3.4 &   4.9 &   24.3 & $2.7 \cdot 10^5$ & $1.0 \cdot 10^5$ &  1.43 \\
NGC  6341 &  1.5 &  1.81 & -8.20 &  1.4 &   9.9 &   25.2 & $5.3 \cdot 10^5$ & $2.0 \cdot 10^5$ &  1.42 \\
NGC  6352 &  0.6 &  1.10 & -6.48 &  3.3 &   3.3 &   16.4 & $1.8 \cdot 10^5$ & $3.7 \cdot 10^4$ &  0.94 \\
NGC  6397 &  1.4 &  2.50 & -6.63 &  3.1 &   6.3 &   18.8 & $1.6 \cdot 10^5$ & $4.5 \cdot 10^4$ &  1.19 \\
NGC  6496$^1$ &  0.7 &  0.70 & -7.23 &  4.3 &   4.3 &   23.1 & $2.2 \cdot 10^5$ & $8.2 \cdot 10^4$ &  1.40 \\
NGC  6656 &  1.4 &  1.31 & -8.50 &  2.9 &   9.3 &   41.2 & $5.5 \cdot 10^5$ & $2.9 \cdot 10^5$ &  1.64 \\
NGC  6712 & -0.9 &  0.90 & -7.50 &  0.9 &   6.2 &   16.4 & $5.1 \cdot 10^5$ & $9.4 \cdot 10^4$ &  0.84 \\
NGC  6752 &  1.6 &  2.50 & -7.73 &  4.8 &   5.6 &   31.8 & $2.9 \cdot 10^5$ & $1.4 \cdot 10^5$ &  1.57 \\
NGC  6809 &  1.3 &  0.76 & -7.55 &  1.9 &   5.8 &   20.9 & $3.3 \cdot 10^5$ & $1.1 \cdot 10^5$ &  1.30 \\
NGC  6838 & -0.2 &  1.15 & -5.60 &  4.5 &   6.7 &   16.2 & $8.4 \cdot 10^4$ & $1.7 \cdot 10^4$ &  0.90 \\
NGC  7078 &  1.9 &  2.50 & -9.17 &  5.4 &  10.3 &   86.0 & $9.0 \cdot 10^5$ & $5.6 \cdot 10^5$ &  1.73 \\
NGC  7099 &  1.4 &  2.50 & -7.43 &  3.0 &   6.9 &   24.6 & $2.6 \cdot 10^5$ & $1.0 \cdot 10^5$ &  1.44 \\
\end{tabular}

\bigskip
{\footnotesize $^1$We assumed a circular orbit for NGC 6496 at its current Galactocentric distance since the 
proper motion of this cluster is not known.}
\end{table*}

\begin{figure}[htbp!]
\plotone{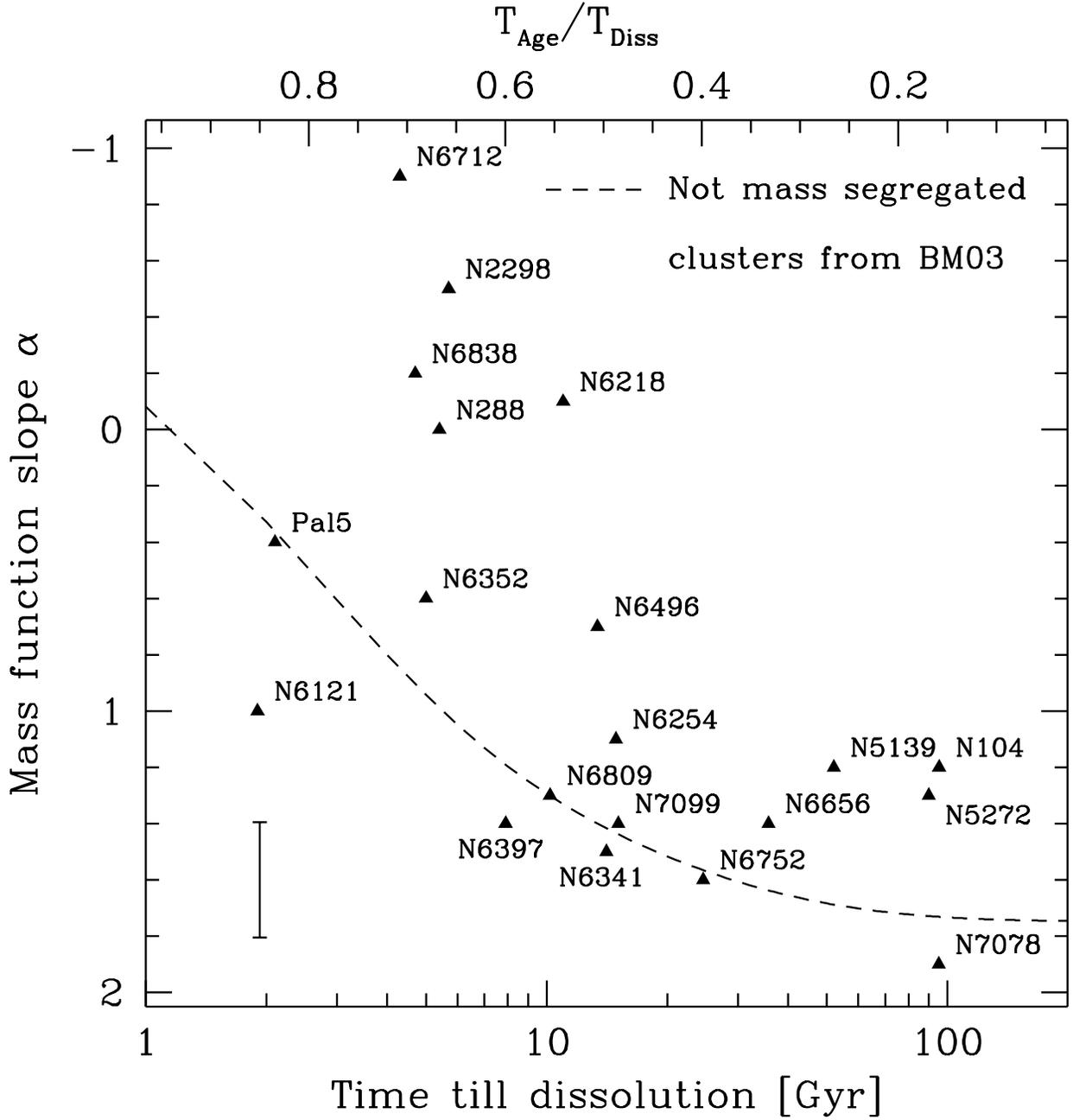}
\caption{Observed mass function slope vs. lifetime remaining to dissolution as determined from the current mass
of each cluster and eqs. \ref{eltime} and \ref{emfunc}.  The errorbar in the lower left corner shows typical uncertainties
of the observed slopes which are around 0.2.  The observed mass function slopes show a clear
correlation in the sense that clusters closer to dissolution are on average more strongly depleted in low-mass stars.
The dashed line shows the expected mass function slope for clusters without primordial mass segregation, 
determined from the models 
of \citet{bm03} (eq.~\ref{mftf}). It provides a good fit for most clusters, however a number of
clusters close to dissolution are much more strongly depleted in low-mass stars.
\label{fig1}}
\end{figure}

Fig.\ \ref{fig1} compares the predicted mass function slopes with the observations. It can be seen that all
clusters with remaining lifetimes larger than $T=20$ Gyr have nearly identical slopes with $\alpha \approx 1.5$. 
This value is close to the expected slope for stars with $0.3 < m < 0.8$ M$_\odot$ drawn from a canonical IMF, 
$\alpha=1.7$. Globular clusters have therefore started 
with a mass function slope at the low mass end which is similar to that seen for open clusters in the present-day universe.
Fig.\ \ref{fig1} also shows that, in agreement with the theoretical results from the $N$-body simulations, the 
mass functions of 
clusters close to dissolution become depleted in low-mass stars. While some clusters lie close to the predictions from the models 
of BM03 (dashed line), a number of clusters are significantly stronger depleted in low-mass stars. In the $N$-body models, 
slopes with $\alpha<0$ are hardly reached since the clusters first have to go into core-collapse to become mass segregated
and then dissolve before reaching a strong enough depletion of low-mass stars. Hence, these clusters cannot be explained 
with the type of initial conditions used by \citet{bm03}, i.e.\ clusters that form in dynamical equilibrium, filling
their tidal radii and start without primordial mass segregation. In addition, the correlation 
of mass function slope and cluster concentration noted by \citet{dpp07} is difficult to understand with 
non-mass-segregated clusters (see Fig. 3, left panel). 

As explained in the Introduction, several lines of evidence
indicate that star clusters form mass segregated, in which case the depletion of low-mass stars could 
happen much quicker than in the models by \citet{bm03}. This offers a possible way how to explain 
the observations. 
We will therefore explore the influence of primordial mass segregation in the next sections. 

\section{$N$-body models of mass segregated clusters}

In order to understand if primordial mass segregation helps reconciling the discrepancy between
observations and simulations, we calculated a number of models starting with primordial mass segregation.
All runs were performed with the collisional $N$-body code NBODY4 \citep{a99} on the GRAPE6 computers
of Bonn University. The modeled clusters contained between $10.000$ to 
$90.000$ stars initially. Since these numbers are rather small compared to particle numbers in globular clusters, 
we decided to omit stellar evolution and start all runs with a power-law mass function with slope $\alpha=1.3$ between
lower and upper mass limits of $m=0.1$ and $m=1.2$ M$_\odot$. This should capture the essential physics of the 
collisional evolution of globular clusters. In order to account for the break in slope of the
canonical IMF at 0.5 M$_\odot$, we assume that for stars more massive than 0.5 M$_\odot$, only a fraction $0.5 M_\odot/m$ of stars 
are main-sequence stars while the other are compact remnants which are not taken into account when mass function slopes 
are determined. All clusters started from King $W_0=3.0$ density profiles
and moved on circular orbits through an isothermal Galaxy. In order to study the influence of the
initial cluster size, we calculated two sets of models, one in which the tidal radius of the external tidal field,
$r_J$, was equal to the tidal radius  of the King model, $r_J/r_t = 1$, and one set of tidally underfilling
models with $r_J/r_t = 3$. The algorithm for creating mass segregated clusters in virial equilibrium is 
described in the Appendix. In our models, we studied the evolution of unsegregated clusters and clusters in which 
the mass and energy arrays are completely ordered before stars are assigned positions and velocities. These models 
therefore show the maximum influence mass segregation can have and realistic clusters should lie between the 
two extremes covered by our simulations. Table~2 summarises the runs performed. 

\begin{table*}[t]
\caption[]{Details of the $N$-body models. The second column gives the initial number of cluster stars,
the third column whether or not primordial mass segregation was present. The fourth column gives
the tidal filling factor $r_J/r_t$ and the last two columns give the dissolution time, defined to be the time
when 99\% of the mass is lost, and the core collapse time, expressed in $N$-body units \citep{hh02}.\\} 
\begin{tabular}{rccccc}
 Nr. & N & PMS & $r_J/r_t$ & $T_{Diss}$ & $T_{CC}$ \\
  & & & & \multicolumn{1}{c}{[NBODY]} & \multicolumn{1}{c}{[NBODY]} \\ 
 1 & 10.0000 & Yes & 1.0 &  $\;\;$1326 & $\;\;$564 \\
 2 & 10.0000 &  No & 1.0 &  $\;\;$1380 & $\;\;$728 \\
 3 & 30.0000 & Yes & 1.0 &  $\;\;$2534 & 1550 \\
 4 & 30.0000 &  No & 1.0 &  $\;\;$2703 & 1686 \\
 5 & 90.0000 & Yes & 1.0 &  $\;\;$5097 & 3582 \\
 6 & 90.0000 &  No & 1.0 &  $\;\;$5666 & 3976 \\
 7 & 10.0000 & Yes & 3.0 &  $\;\;$6522 & $\;\;$778 \\
 8 & 10.0000 &  No & 3.0 &  $\;\;$6160 & $\;\;$886 \\
 9 & 30.0000 & Yes & 3.0 & 11850 & 1970 \\
10 & 30.0000 &  No & 3.0 & 11960 & 2130 \\
\end{tabular}
\end{table*}

\section{Results for mass segregated clusters}

\begin{figure}[htbp!]
\plotone{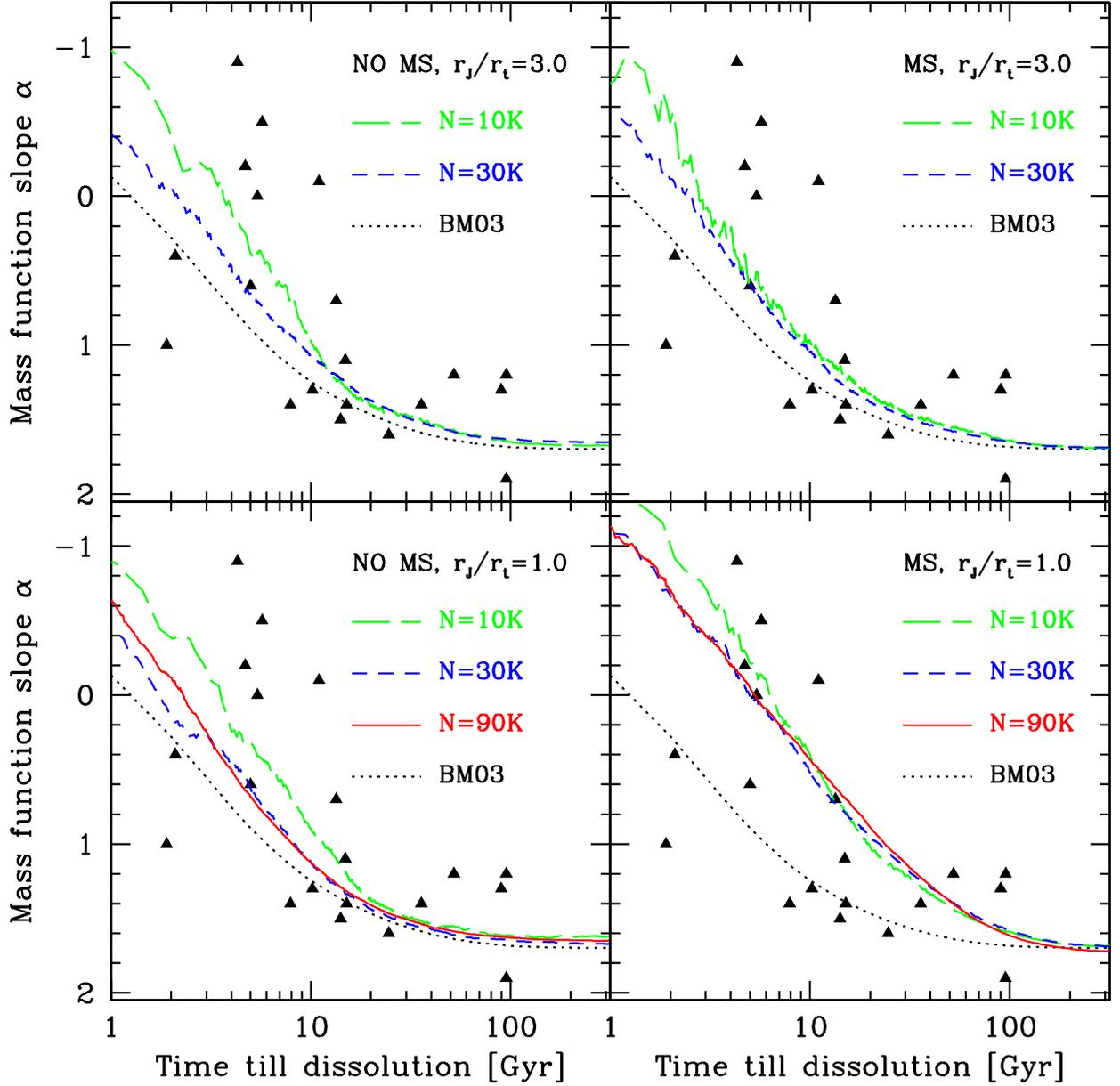}
\caption{Same as Fig.\ 1, but now for mass segregated and not mass segregated clusters. The upper panels depict
the evolution of tidally underfilling clusters with $r_J/r_t=3.0$, the lower panels that of clusters with $r_J/r_t=1.0$. 
The left panels depict the evolution of non-segregated clusters, the right panels that of mass segregated clusters.
Mass segregated clusters in strong tidal fields lose low-mass stars  
right from the start of the simulations, leading to more strongly depleted mass functions by the time the clusters
are close to dissolution. This can explain the mass function slopes of strongly depleted globular clusters. 
In most cases, mass functions evolve nearly independently of the initial number of stars.
\label{fig2}}
\epsscale{1.0}
\end{figure}

Fig.\ \ref{fig2} depicts the evolution of the mass function of initially mass segregated clusters
starting from various initial conditions and compares it with the evolution of non-segregated clusters
and observations of Galactic globular clusters. Final mass functions were determined from a fit to
the distribution of stars in the mass range $0.3 M_\odot < m < 0.8 M_\odot$, similar to the mass range
for which observed mass functions are determined for most star clusters. It can be seen that in tidally underfilling models (upper panels
with $r_J/r_t=3.0$), the evolution does not depend much on whether the cluster initially starts mass segregated
or not. This is probably due to the short core collapse times of strongly concentrated clusters
compared with their dissolution times (see Table~2). Since the starting condition has largely been erased by the time
a cluster goes into core collapse, and since the pre-core collapse evolution typically lasts only 
about 20\% of the
total lifetime for these models, the starting condition should not strongly influence the overall evolution.
The lower panels in Fig.\ \ref{fig2} depict the evolution of tidally filling clusters with $r_J/r_t=1.0$.
While non-segregated clusters still evolve close to the prediction of \citet{bm03} (dotted lines), the evolution of
mass segregated clusters is now markedly different: Since in mass segregated clusters, low-mass stars start
close to the tidal radius, they are being depleted right from the start of the simulations, leading to final mass
functions much more strongly depleted in low-mass stars. The amount of depletion is strong enough to explain
most observed mass functions. Hence, the range of slopes seen for Galactic globular clusters can, at least in
principle, be explained if some started mass segregated while others didn't, or all of them started mass segregated but 
with a range of tidal filling factors.  We also note that the expulsion of residual gas within the first Myr can enhance
the depletion of low-mass stars if the clusters start mass segregated \citep{mkb08}.

\begin{figure}[tbp!]
\plotone{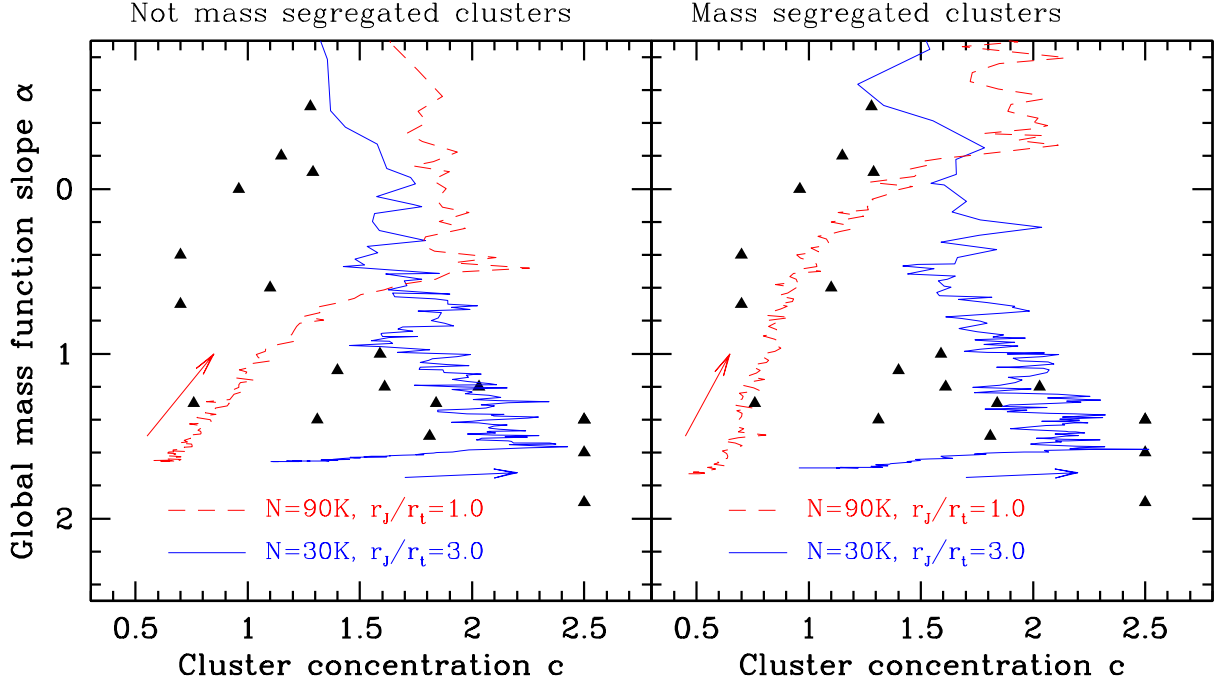}
\caption{Evolution of initially non-mass-segregated (left panel) and mass segregated (right panel) clusters
in the concentration vs. mass function slope plane. The arrows mark the direction in which the clusters are evolving. 
The cluster concentration first increases as the clusters go into 
core collapse and then decreases again in post-collapse evolution. Initially not mass segregated clusters are always
in post-collapse by the time they have become significantly depleted in low-mass stars, making it impossible
to explain clusters with both low concentration and strongly depleted mass functions. In contrast, clusters 
filling their tidal boundary ($r_J/r_t=1.0$) and with primordial mass segregation become significantly 
depleted in low-mass stars before going into core collapse. \label{fig3}}
\end{figure}

Fig.~\ref{fig3} finally depicts the evolution of star clusters in the concentration vs. mass function slope plane.
In order to determine the concentration, King models were fitted to the surface density distribution of stars with 
masses in the range
$0.6 \le m \le 0.8$ M$_\odot$ and the concentration $c$ was chosen from the King model which gave the best fit to 
the simulated clusters.
The restriction to stars in the mass range $0.6 \le m \le 0.8$ M$_\odot$ was done since in globular clusters these would be 
the stars which create most of the cluster light.
As can be seen, the cluster concentration first increases in all models as the clusters go into core collapse and then decreases again in 
post-collapse due to core expansion driven by binaries in the cluster centre. In post-collapse, all models reach a stable 
value of $c=1.6$ nearly independent of the initial concentration. Clusters also move upward 
in Fig.~\ref{fig3} as the mass function becomes depleted in low-mass stars.
For initially non-segregated clusters (left panel), core collapse is fast enough that they are always
in post-collapse by the time they have become strongly depleted in low-mass stars. Especially concentrated
clusters hardly lose any stars in the pre-collapse phase. Since our clusters
started from already very low-concentration, King $W_0=3.0$ models, it seems impossible to delay core collapse 
much further by doing simulations of even lower concentration models. Hence, as was already noted by \citet{dpp07}, 
one cannot explain low-concentration clusters which are strongly depleted in low-mass stars by non-segregated models
assuming that the IMF of stars is universal.

The right panel of Fig.~\ref{fig3} depicts the evolution of initially mass segregated clusters. For tidally underfilling
clusters with $r_J/r_t=3.0$, the evolution is virtually indistinguishable from the evolution of non-segregated clusters
with $r_J/r_t=3.0$. Clusters with $r_J/r_t=1.0$ on the other hand lose low-mass stars much quicker and go into core collapse
only after their mass function has become significantly depleted in low-mass stars. Primordial mass segregation would
therefore also provide an explanation for low concentration globular clusters which are strongly depleted in low-mass stars.

\section{Conclusions}

We have followed the dynamical
evolution of star clusters in tidal fields starting with and without primordial mass segregation. We
find that clusters with primordial mass segregation lose their low-mass stars more rapidly than non-segregated 
ones if being immersed in a strong external tidal field, due to the fact that low-mass stars start their
lifes in the outer cluster parts where they can easily be removed by the tidal field. For 
clusters in weaker tidal fields, primordial mass segregation makes only a small difference 
to the cluster evolution since strong mass loss starts only after core collapse, by which 
time cluster evolution has largely erased the initial conditions.
For all studied models, the absolute values of the core collapse time and the lifetime 
decrease by no more than 10\% due to the introduction of primordial mass segregation. The difference
could be larger for simulations which also include the effects of stellar evolution, although
e.g.\ \citet{aetal08} found only a slight increase in core-collapse time for mass segregated clusters
compared to non-segregated ones.

Our simulations show that primordial mass segregation is a way to explain the strong depletion of 
low-mass stars seen in some globular clusters as well as the correlation between mass 
function slope and cluster concentration recently found by \citet{dpp07}. Given the strong
observational evidence for primordial mass segregation in young star clusters, we conclude
that at least some, but possibly all, globular clusters started mass segregated. 
The range of mass function slopes seen for Galactic globular clusters can then be explained 
if they started with a range of tidal filling factors but all of them had the same initial mass 
function slope. Also, the clusters in the \cite{dpp07} sample span a range of metallicities
$-2.2 < \mbox{[Fe/H]} < -0.6$ and formed at high redshifts, while current ($z \sim 0$) star formation
with $\mbox{[Fe/H]} \approx 0.0$ produces an indistinguishable IMF.
Our results therefore indicate that the initial mass function of low-mass stars has been more or less 
universal for a large range of star formation environments, redshifts and cluster metallicities.

The effect of primordial mass segregation on the mass function is enhanced if 
residual gas removal is taken into account, since due to the sudden drop of the cluster potential as 
a result of gas expulsion, stars at large radii are preferentially lost from star clusters. Gas expulsion
also naturally leads to tidally filling clusters. The influence of this effect together with the effect of 
unresolved binaries on the observed mass functions is discussed in \citet{mkb08}. Their study shows
that the effect of gas expulsion depends on several parameters, like the amount of gas removed 
(i.e.\ the star-formation efficiency), the timescale over which gas expulsion happens and how strongly 
the proto-globular cluster is immersed in an external tidal field (see the grid
of models run by Baumgardt \& Kroupa 2007). Due to their high masses, embedded globular clusters
must have started with ratios of half-mass radius to tidal radius, $r_h/r_t$, significantly smaller 
than 0.1. Also, the crossing time of a $r_h=1$ pc, $10^6$ M$_\odot$ proto-globular
cluster is only 20.000 yr, while e.g.\ \citet{bkp08} found that gas expulsion from globular clusters
should take several $10^5$ to $10^6$ yr. Hence the primordial gas was probably removed adiabatically (i.e. on a 
timescale much longer than the crossing time of the cluster) from globular clusters.
As can be seen from fig.\ 3 of Marks et al. (2008), clusters with $r_h/r_t$ values smaller than 0.06 
and adiabatic gas removal mostly preserve their IMF or receive only small changes to it, even
when ending up with low concentrations. Hence, while primordial gas expulsion might
contribute to the change in the IMF, gas expulsion alone is not likely to explain strongly depleted
mass functions in globular clusters.

Primordial binaries also influence measured mass function slopes because a fraction of low-mass stars is hidden
in binaries with more massive primaries and because cluster evolution, especially the 
evolution after core-collapse, is different if primordial binaries are present. The influence of hidden
low-mass stars on the mass function slope is also discussed in \citet{mkb08}. The influence of primordial binaries on cluster evolution is
less clear since for example the simulations by \citet{fr07} show that clusters with primordial binaries 
reach concentrations around $c \approx 2$ in the post-collapse phase, which is close to the values found 
here for clusters without primordial binaries. 
Also, in mass segregated, multi-mass clusters, primordial binaries are likely to have a smaller effect on 
the evolution, 
since the cluster evolution is driven by only few active binaries. If massive stars start their life in 
the core, they quickly form binaries and the later cluster evolution becomes indistinguishable 
from clusters with primordial binaries.

It therefore remains to be seen how results change for models which
self-consistently include the effects of gas expulsion, two-body relaxation and primordial binaries.
We plan to carry out such studies in the future.

We finally suggest a new method for setting-up mass segregated clusters, which has the advantage that it 
always creates clusters which are in virial equilibrium since the mass density profile is not changed due to the introduction of 
mass segregation. It is also flexible and can work with any given mass density profile, initial mass function of
stars and can be combined with any scheme for setting up mass segregation.

\section*{Acknowledgments}

The authors would like to thank Sverre Aarseth for his constant help with the NBODY4 code and Eliani Ardi
for useful discussions.

\appendix

\section{Creation of mass segregated clusters}

Primordial mass segregation is introduced by first creating a set of $N'$ positions and velocities for an unsegregated
cluster, distributed according to the desired mass density profile (King profiles in our case). This set is then
ordered according to the specific energy of each star (potential plus kinetic), putting lowest-energy stars first.
In a second step, we create an array of $N$ masses, distributed according to the desired mass function, and sort
this array in descending order. Here $N$ is the number of stars in the final cluster. We then calculate the 
cumulative mass function $M_{Cum}[i] = \sum_{i'=1}^i M[i']$ of all stars in the mass array and divide $M_{Cum}[i]$
by the total mass, so that $M_{Cum}[i]$ runs from 0 to 1. We finally pick a position and velocity for each star $i$ in 
the mass array by randomly choosing an entry between $N' M_{Cum}[i-1]$ and $N' M_{Cum}[i]$ from the energy array. 
In order to make sure that there 
is at least one entry from which to choose a position and velocity, the energy array has to contain at least 
$N' = N <\!m\!>/m_{Low}$ stars, where $<\!m\!>$ and $m_{Low}$ are the average mass of stars and the mass of the 
lowest mass star in the mass array.

\begin{figure}[tbp!]
\plotone{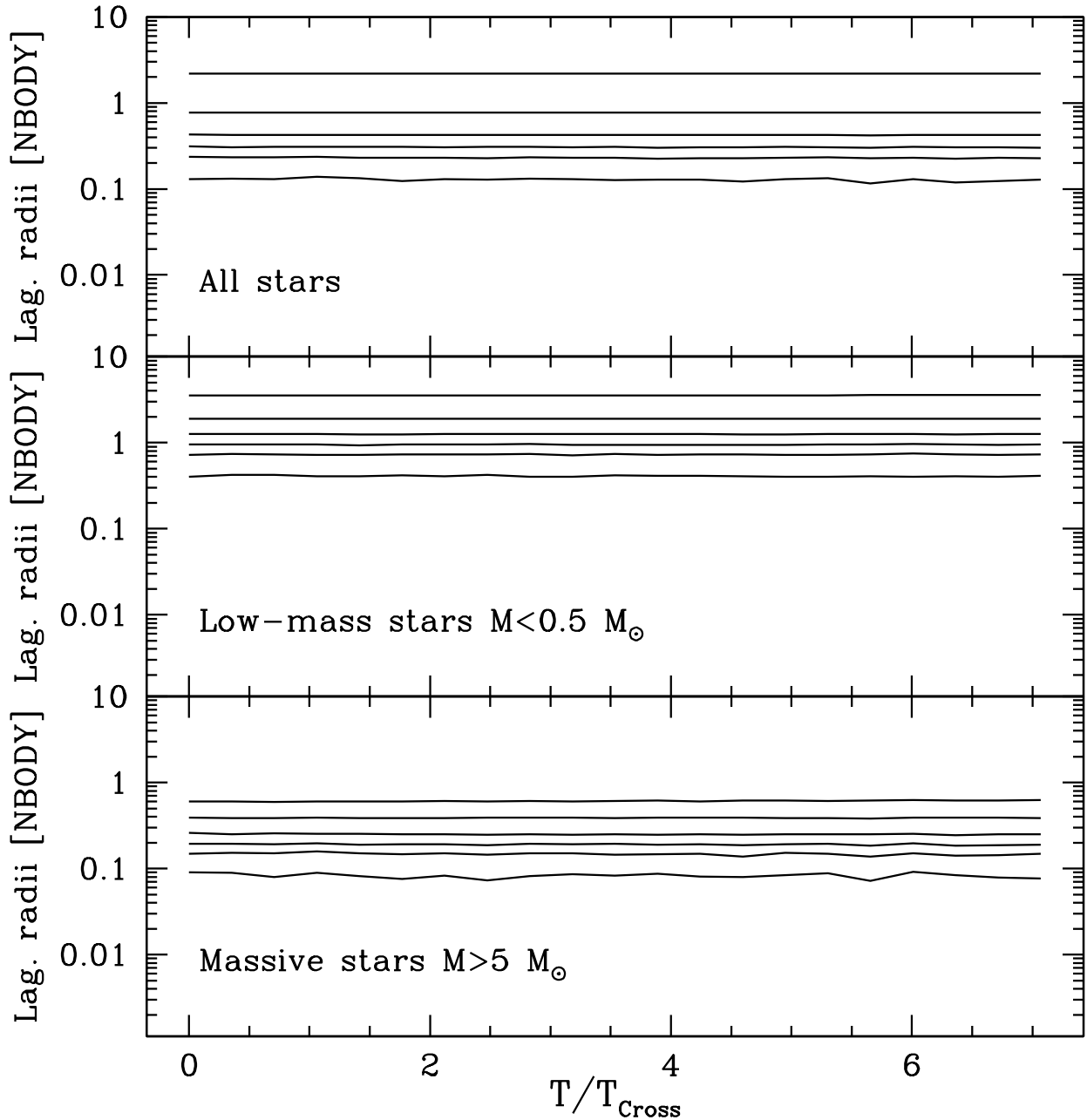}
\caption{Evolution of Lagrangian radii for a completely mass-segregated Plummer model as a function of time
(which is expressed in units of initial crossing time). Shown is the evolution of Lag.\ radii 
containing (from bottom to top) 1\%, 5\%, 10\%, 20\%, 50\% and 90\% of the total mass
of each component. As can be seen,  not only are the Lagrangian radii for all stars stable, but also those of
each individual sub-component. 
\label{fig4}}
\end{figure}

The above method has the advantage that it creates clusters which are in virial equilibrium since the mass   
density profile is not changed due to the introduction of mass segregation. In contrast to a method which
sorts stars according to radii rather than energies, it also creates clusters in which each individual mass 
group is in virial equilibrium (see discussion in \citet{aetal08}). The above method is also fast since
the most time consuming
part of the calculation, the sorting of the energies, is only of order $O(N' log N')$. By introducing only partial ordering
in the mass or energy array one can create clusters with a smaller amount of mass segregation. Fig.\ \ref{fig4}
shows as an example the evolution of Lagrangian radii for a Plummer model 
with a Salpeter like mass function going from $0.1 M_\odot$ to $15 M_\odot$ which is 100\%
mass sorted initially As can be seen, not only are the Lagrangian radii of all stars stable, but
also those of individual mass groups. We finally note that a different method for creating mass segregated 
clusters, which uses mean interparticle potentials, has recently also 
been suggested by \citet{setal08}. Compared to their method, the method suggested here has the advantage
that it is more straightforward to create clusters with a desired density profile and that our clusters are always in virial
equilibrium.

\end{document}